\documentclass{icrc2009}

\usepackage{graphicx}   
\usepackage[caption=false]{caption}    
\usepackage[font=footnotesize]{subfig} 
\usepackage{fixltx2e}
\usepackage{url}

\newcommand{\shorttitle}[1]%
{\markboth{Proceedings of the 31\MakeLowercase{$^{st}$} ICRC, {\L}\'{o}d\'{z} 2009}{#1} }

\begin{document}
\title{Reconstruction of Atmospheric Neutrinos in Antares}

\author{\IEEEauthorblockN{Aart Heijboer\IEEEauthorrefmark{1}, for the Antares Collaboration}
\\
\IEEEauthorblockA{\IEEEauthorrefmark{1}Nikhef, Amsterdam.}}

\shorttitle{Aart Heijboer, Neutrino Reconstruction in Antares}
\maketitle

\begin{abstract}

In May 2008, the Antares neutrino telescope was
completed at 2.5 km depth in the Mediterranean Sea;
data taking has been going on since. A prerequisite
for neutrino astronomy is an accurate reconstruction
of the neutrino events, as well as a detailed
understanding of the atmospheric muon and neutrino
backgrounds. Several methods have been developed to
confront the challenges of muon reconstruction in
the sea water environment, which are posed by e.g.
backgrounds due to radioactivity and bioluminescence.
I will discuss the techniques that allowed Antares
to confidently identify its first neutrino events, as
well as recent results on the measurement of atmospheric
neutrinos. 
\end{abstract}


\section{Introduction}

 The Antares collaboration is currently operating
 a 12-line neutrino telescope in the 
 Mediterranean Sea, at a depth of about 2500 m, 40 km from 
 the shore of southern France (see \cite{pas} for 
 a full status report). The full detector
 comprises 12 lines, each fitting 75 Optical
 Modules (OMs), arranged in 25 triplets, also called
 `floors', which are placed 14.5 metres apart along 
 the line. The OMs house a 10 inch photomultiplier
 tube, which is oriented downward at a 45 degree 
 angle to optimise the detection efficiency for 
 neutrino-induced, upgoing muons. The positions
 of the OMs are measured using the systems described
 in \cite{anthony}.

 The deployment of the first detector line took place
 at the beginning of 2006. This line was used to measure 
 the flux of atmospheric muons using a specialised
 reconstruction algorithm \cite{line1paper}. 
 By January 2007, five such detector lines were 
 operational, allowing the application of the methods 
 developed for the 3d reconstruction of muon 
 trajectories. This led to the identification of the 
 first neutrino events; see Fig. \ref{fig_evt}.
 The 12th line was deployed in May 2008, which 
 completed the construction of the detector. 
 
\section{Muon Reconstruction}

 The challenge of measuring muon neutrinos consists 
 of fitting the trajectory of the muon to the arrival
 times, and -to a lesser extent- to the amplitudes
 the Cherenkov light detected by the OMs.
 
\begin{figure}[!ht]
  \centering
  \includegraphics[width=3.0 in]{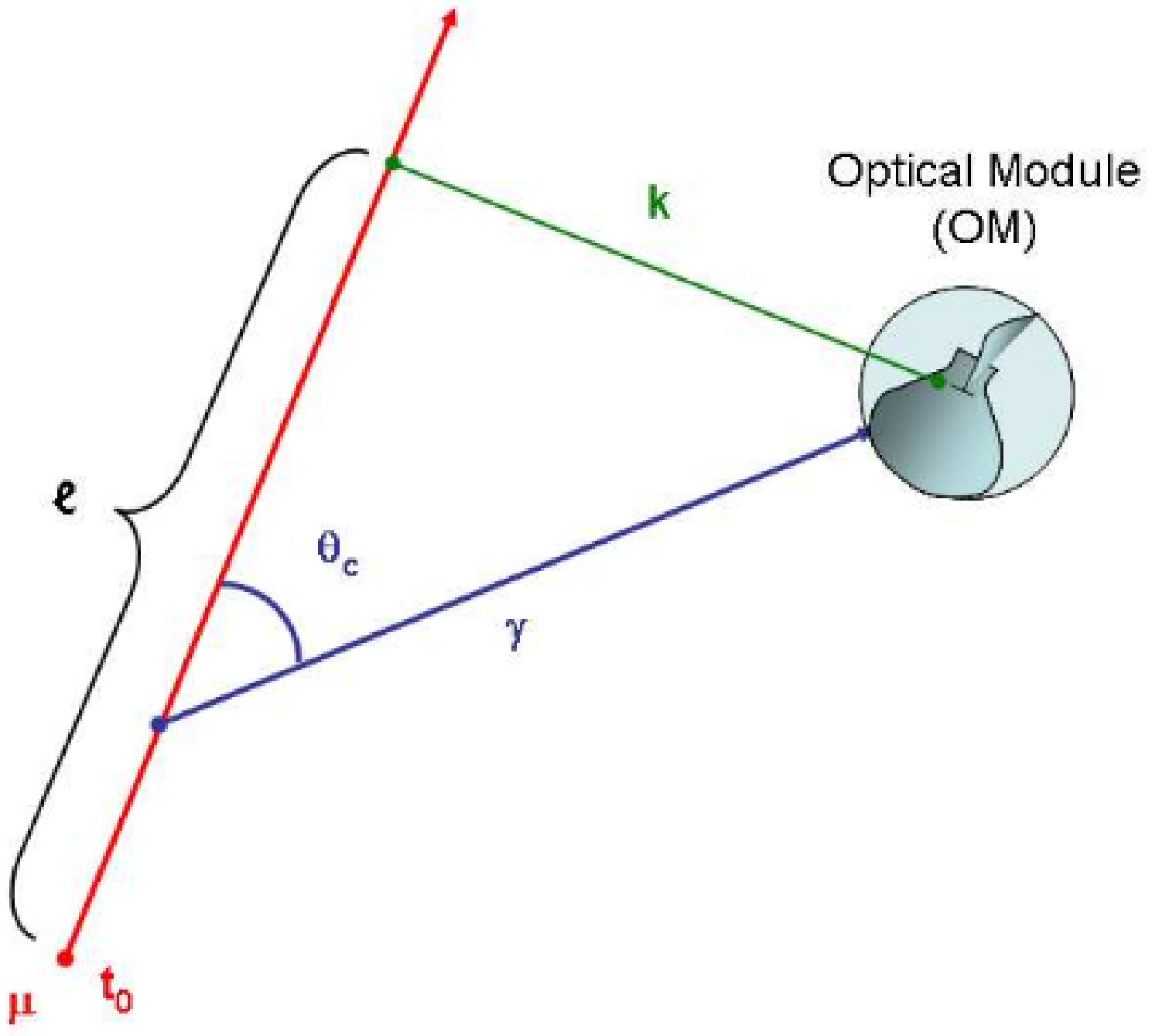}
  \caption{Schematic representation of the relation between the
           muon trajectory and the OM. The line labelled $\gamma$
           indicates the path travelled by a Cherenkov photon from 
           the muon
           to the OM.}
  \label{fig_schema}
\end{figure}

 For a given muon position (at an arbitrarily chosen
 time $t^0$) and direction, the 
 expected arrival time $t^{\rm exp}$ of 
 the Cherenkov photons follows from the geometric
 orientation of the OM with respect to the muon path:
\begin{equation}
t^{\rm exp} = t^0 + {1 \over c } \bigl( l - {k \over {\tan \theta_C}} \bigr) +
                    {1 \over v } \bigl(     {k \over {\sin \theta_C}} \bigr),
\end{equation}
 where the distances $l$ and $k$ are defined in Fig. \ref{fig_schema},
 $\theta_C$ is the Cherenkov angle ($\sim 42^o$ in water) and $v$ is
 the group velocity of light in the water.
 The difference between
 $t^{\rm exp}$ and the measured arrival time of 
 the photon (i.e. the 'hit time) defines the time 
 residual: $r \equiv t^{\rm measured} - t^{\rm exp}$.

\begin{figure}[!t]
  \centering
  \includegraphics[width=3.5in]{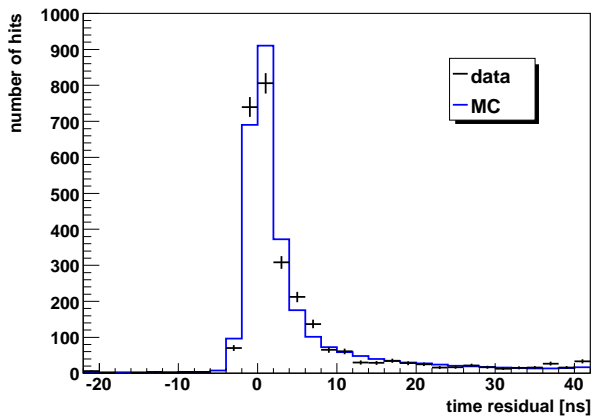}
  \caption{Time residuals of the hits with respect to the result
           of the full likelihood fit for selected, upgoing events
           (neutrino candidates). The data were taken in 2007 
            with 5 detector lines. The peak shows the (order 1 ns) intrinsic
            timing resolution of the OMs. The tail is due to light
            from secondary particles and to scattered photons.}
  \label{fig_res}
\end{figure}

\begin{figure}[!ht]
  \centering
  \includegraphics[width=3.0in]{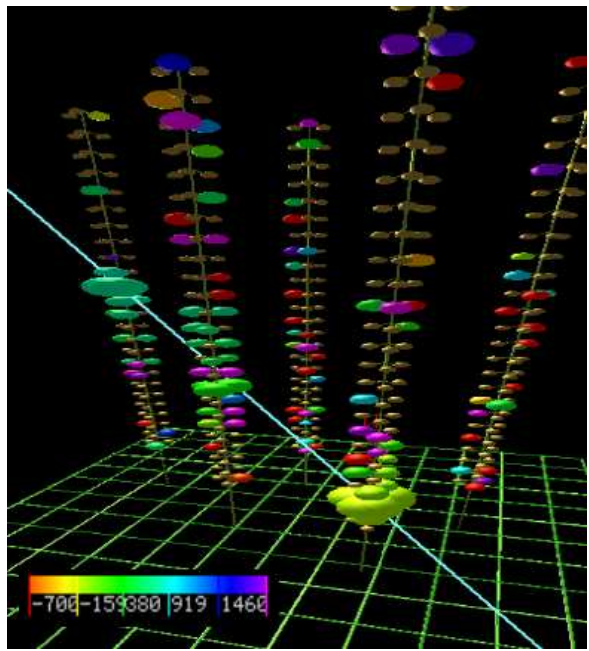}
  \caption{Event display of one of the first neutrino 
           candidates detected with Antares. This event was
           found and reconstructed using the full 
           likelihood fit. The colour of the hits indicates
           the time (according to the legend in the lower left
           corner), while the size indicates the collected charge.}
  \label{fig_evt}
\end{figure}

 Photons that scatter in the water and photons
 emitted by secondary particles 
 (e.g. electromagnetic showers created 
 along the muon trajectory) will arrive at the OM
 later than $t^{\rm exp}$, leading to positive 
 residuals. The residual distribution obtained 
 from in data is shown in Fig. \ref{fig_res}. 
 The tail on right due to late photons is 
 clearly visible. 

 The reconstruction algorithms attempt to find muon
 track parameters (i.e. three numbers for the position and two 
 for the direction) for which the residuals are small. This
 can be done by minimising a quantity like $\chi^2 = \sum_{i=1}^{N_{\rm hits}} r_i^2$
 or by maximising the likelihood function $\log L = \sum_{i=1}^{N_{\rm hits}} \log P(r_i)$. 
 Here, $P(r)$ is the probability density function (PDF) for the residuals.
 While relatively simple, the equation to compute the residuals
 is non-linear in the track parameters. As a consequence, iterative 
 methods are required for minimising a $\chi^2$-like 
 variable or maximising the likelihood.

 A complicating factor in the reconstruction process 
 is the presence of
 background hits, caused by decaying $K^{40}$ in the sea 
 water and by aquatic life (bioluminescence). If not
 accounted for in the muon reconstruction, the background
 light degrades both the angular resolution and the
 reconstruction efficiency. Antares has developed several
 strategies to deal with this problem. Two of them
 will be discussed in the following sections. We refer
 to these two as the 'full likelihood' and 'online' algorithms.
 They are currently widely used for data analysis. 

\section{Full Likelihood Fit}

 The first algorithm was developed 
 several years before deployment of the detector and
 is described in detail in \cite{aartstrat}.
 This method is based on a likelihood fit that uses a
 detailed parametrisation, derived from simulation,
 for the PDF of the arrival time of the hits $P(r)$, which takes
 into account the probability of hits arriving
 late due to Cherenkov emission by secondary particles
 or light scattering. Moreover, the probability of a hit
 being due to background is accounted for as a function 
 of the hit amplitude and the orientation of the OM with 
 respect to the muon track. 
 It was found that the likelihood function has many local 
 maxima and that the likelihood fit is only successful if the
 maximisation procedure is started with 
 track parameters that are already a good approximation
 to the optimal solution. To obtain this approximate
 solution, the full likelihood fit is preceded by 
 a series of `prefit' algorithms of increasing sophistication.
 An important ingredient in the prefit stage is the use 
 of a so-called `M-estimator', which is a variant of a
 $\chi^2$-fit in which hits with large residuals are
 given less importance compared to a regular $\chi^2$. 
 This is crucial, as it allows the fit to converge
 to a solution relatively close (typically a few degrees) to 
 the true muon parameters, while being robust against the 
 presence of background hits at large residuals. 
 The M-estimate is followed by
 two different versions of the likelihood fit, the last
 of which fully accounts for the presence of background
 hits. The procedure is started at nine different starting 
 points to increase the probability of finding the global 
 minimum. To mitigate the associated loss in speed, 
 analytical expressions for the gradient of the 
 likelihood function are used in the min/maximisation 
 processes.

 The value of the final log-likelihood per degree
 of freedom that is obtained from the final fit
 is used as a measure of the goodness of fit. 
 This is combined with information on the number
 of times the repeated procedure converged to the
 same result, $N_{\rm comp}$ to provide a value 
 $\Lambda=\log(L)/N_{\rm dof} - 0.1(N_{\rm comp} -1 )$.
 The variable $\Lambda$ can be used to reject badly 
 reconstructed events; in particular atmospheric 
 muons that are reconstructed as upward-going. An
 example of the use of this algorithm for reconstructing
 and selecting neutrinos for a point source search is
 given in \cite{point}.

\subsection{Results}

 The full likelihood fit is optimised for the high energy
 neutrinos that are expected from astrophysical sources 
 ($E_{\nu}^{-2}$ spectrum, yielding muons in the 
 multi-TeV range). Simulations indicate that, 
 with this algorithm, Antares reaches an angular 
 resolution (defined as the median angle between the 
 true and reconstructed muon) smaller than 0.3 
 degrees for neutrino energies above 10 TeV. Below 
 this energy the scattering angle of the neutrino 
 interaction dominates.

 As the hit residuals are the main ingredient driving the
 angular resolution, the agreement observed between data
 and simulation in the residual distribution (see Fig. 2) 
 is good evidence that the algorithm is performing as 
 expected.

\section{Online Algorithm}

 The second approach to reconstructing tracks in the
 presence of background hits was developed 
 during the commissioning of the detector for the
 online event display that is used for monitoring the detector. We
 therefore refer to this fit as the `online algorithm', although
 it is now also frequently used for offline data analysis.

 Whereas the full likelihood fit described above attempts to incorporate
 all the background hits in the fit, the philosophy of the online
 algorithm is to select a very high purity sample
 of signal hits. This is followed by a fit of the muon 
 trajectory using a model that can be relatively simple. In this
 case a $\chi^2$-like fit is performed.

 The algorithm merges hits on the three OMs in a floor and 
 uses the centre of the triplets in the fit. While degrading
 the timing precision (and therefore, in theory the angular
 resolution) somewhat, this does make the algorithm independent
 of measurements of the azimuthal orientation of the triplets, which
 vary due to sea currents and which normally 
 need to be measured using compasses located on the floors.

\subsection{Hit Selection}

 The selection of hits starts by identifying floors
 that collected an amplitude (i.e. charge collected on the PMTs) 
 corresponding  to more than  2.5 photo-electrons (pe), 
 or 1.5 pe in case multiple hits 
 were detected within a time window 
 of 20 ns. Such configurations are rarely produced by optical
 backgrounds, but occur in most of the signal events.
 Doublets of such floors are identified, allowing for no
 more than one empty floor in between and requiring that
 the time difference of the hits is smaller 
 than $80 {\rm ns}$ per floor of separation. Only the 
 detector lines which at least one such doublet are 
 used in the fit. 
 Clusters of hits are then formed by iteratively 
 complementing the 
 doublets with adjacent hits that
 are close in time (within $80 {\rm ns}$ per floor) and 
 distance (no more than one empty floor) to 
 the already identified cluster.

\subsection{Fit}

 The fit is performed using a score function that is derived
 from the expression for the $\chi^2$ of the hit residuals, with
 an added term that promotes solutions where hits with large
 amplitudes pass the OM at close range. The minimised function is:
\begin{equation}
Q = \sum_i \bigr[ {1 \over {\sigma^2}} r_i^2 + \alpha q_i d_i \bigl],
\end{equation}
 where the sum is over all selected hits and where 
 $r_i$ is the residual of hit $i$, $q_i$ is the amplitude associated
 with that hit, and $d_i$ is
 the distance from the hypothesised track to the OM. The constants 
 $\sigma$ and $\alpha$ were set to $10 \rm ~ns$ and $50 ~\rm m^{-1} pe^{-1}$.

 For events with multiple selected lines, the position of
 the track is determined using the $Q$ fit to 
 minimise the hit residuals. An additional fit is performed
 using a `bright point' hypothesis corresponding to a single,
 localised flash of light. A comparison of the
 quality of the two fits is used to reject 
 events in which downgoing muons create a bright 
 electromagnetic shower that may mimic an upgoing track.

 Events with only one selected detector line do not carry information
 on the azimuth direction of the muon. Hence, for these events,
 a four-parameter fit is performed, yielding the zenith angle
 of the muon track. Also in this case, a bright point fit is
 done for comparison.


 As in the full likelihood fit, the value of $Q$ found by the fit is used to reject badly 
 reconstructed events;
 in particular atmospheric muons that are reconstructed 
 as upward-going. The track fit quality is required to be
 better (i.e. smaller) than 1.35 (1.5) for reconstruction
 with two (more than two) lines; Events with a bright-point
 fit quality better than 1.8 are vetoed.

 The fact that only a single minimisation is performed 
 makes the online algorithm about an order of magnitude
 faster than the algorithm used for the full fit, which
 typically performs 20 full minimisations.

 With the current trigger setup, both algorithms are 
 fast enough to run on all triggered events in real 
 time on a single CPU.

\subsection{Results}

 Simulations show that, at high energies, the online algorithm 
 achieves a typical angular resolution on the muon direction 
 of 2 degrees (1 degree for more stringent cuts), independently 
 of the energy. At high energies, 
 this is a factor 6 to 3 significantly worse than the resolution o
 f the full likelihood fit, which is not unexpected as the assumption of 
 Gaussianity of the time residuals that underlies the $\chi^2$-fit
 is known to be an approximation. The online algorithm is therefore
 not well suited for those neutrino astronomy studies that require 
 optimal angular resolution. On the other hand, the simplicity 
 of the algorithm and the (expected) robustness against inaccuracies in the 
 detector description, have made it a good alternative for the 
 initial studies of the atmospheric muon (see \cite{atm_mu}) and
 neutrino fluxes.  

\begin{figure}[!t]
  \centering
  \includegraphics[width=3.3in]{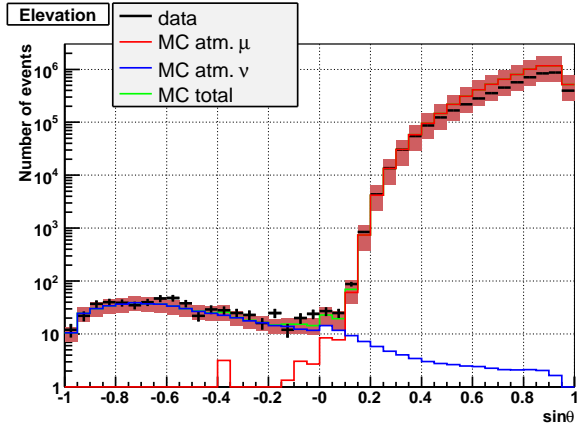}
  \caption{Distribution of the sine of the elevation angle for muons obtained             
           from the multi-line online reconstruction algorithm 
           for the 2008 data, i.e. 9-12 lines.
           }
  \label{fig_bbm}
 \end{figure}

 Figure \ref{fig_bbm} shows the elevation above
 the horizon for the data taken in 2008 for the 
 multi-line events in comparison to simulation. 
 The data were taken using a 9,10 and 12 line detector 
 and represent
 a total live time of 173 days.
 The atmospheric muons were simulated using Corsika
 with Horandel fluxes and the QGSJET hadronic model. 
 A combined theoretical and systematic uncertainty
 of 30 (50)\% on the expected number of 
 neutrinos (muons) accounts for uncertainties in the 
 (primary) flux and interaction model, and in detector 
 acceptance. 
 
 A total of 582 upward going multi-line events were reconstructed, whereas
 the simulation predicts 494(13) events due to atmospheric 
 neutrinos (muons). The difference, a factor of 0.87, is
 within the systematic  uncertainty on the simulation. 

\begin{figure}[!t]
  \centering
  \includegraphics[width=3.3in]{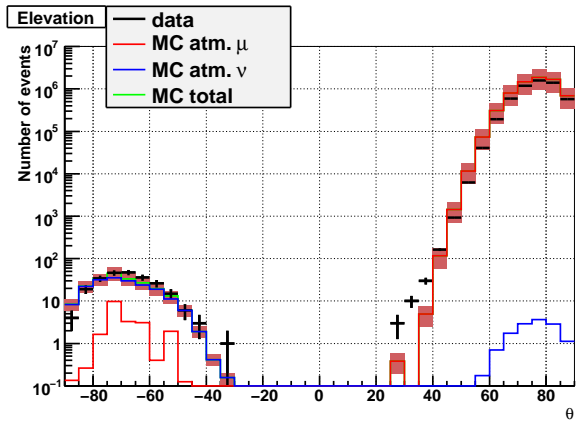}
  \caption{Distribution of the elevation angle f muons reconstructed
           on a single line using the online reconstruction algorithm for 2008 
           data, i.e. 9-12 lines.
           }
  \label{fig_bb1}
 \end{figure}

\vspace*{10cm}
\newpage

 Figure \ref{fig_bb1} shows the elevation angle for events
 fit on a single line using the online algorithm. The efficiency
 for such events is greatly enhanced for vertical (either upward, 
 or downward-going) tracks. The data agree well with the simulation,
 which is dominated by the atmospheric neutrino
 contribution for upward-going tracks. A total of 237 upward going 
 single-line events are found, while simulation predicts 190 (21) due to
 atmospheric neutrino (muon) events. The MC/data ratio is 0.89, which 
 is similar to the ratio reported above for multi-line events.

\section{Conclusion} 

 Since the deployment of the first five lines of the detector, 
 the Antares collaboration has been routinely detecting muons
 and atmospheric neutrinos. About five high-quality upgoing
 neutrino candidates are detected per day. The number of detected 
 neutrinos and their zenith and azimuth angle distributions
 agree well with simulations. 
 The observed time residuals of the hits with respect to reconstructed
 neutrino candidate tracks is in good agreement to the corresponding
 simulation. This further strengthens
 the confidence that Antares is performing as expected
 and that it is achieving sub-degree angular resolution.

\end{document}